# 15. PLANNING OPTIMAL FROM THE FIRM VALUE CREATION PERSPECTIVE. LEVELS OF OPERATING CASH INVESTMENTS

Grzegorz MICHALSKI[*]

## Abstract

*The basic financial purpose of corporation is creation of its value. Liquidity management should also contribute to realization of this fundamental aim. Many of the current asset management models that are found in financial management literature assume book profit maximization as the basic financial purpose. These book profit-based models could be lacking in what relates to another aim (i.e., maximization of enterprise value). The corporate value creation strategy is executed with a focus on risk and uncertainty. Firms hold cash for a variety of reasons. Generally, cash balances held in a firm can be called considered, precautionary, speculative, transactional and intentional. The first are the result of management anxieties. Managers fear the negative part of the risk and hold cash to hedge against it. Second, cash balances are held to use chances that are created by the positive part of the risk equation. Next, cash balances are the result of the operating needs of the firm. In this article, we analyze the relation between these types of cash balances and risk. This article presents the discussion about relations between firm's net working investment policy and as result operating cash balances and firm's value. This article also contains propositions for marking levels of precautionary cash balances and speculative cash balances. Application of these propositions should help managers to make better decisions to maximize the value of a firm.*

[*] *Wroclaw University of Economics, Department of Corporate Finance and Value Management, ul. Komandorska 118/120, pok. 1-Z (KFPiZW), PL53-345 Wroclaw, Poland, Grzegorz.Michalski@ue.wroc.pl; http://michalskig.ue.wroc.pl/*

## 1. Introduction

Corporate cash management depends on demands for cash in a firm. The aim of cash management is such that limiting cash levels in the firm maximizes owner wealth. Cash levels must be maintained so as to optimize the balance between costs of holding cash and the costs of insufficient cash. The type and the size of these costs are partly specific to the financial strategy of the firm.

In addition, cash management influences firm value, because its cash investment levels entail the rise of alternative costs, which are affected by net working capital levels. Both the rise and fall of net working capital levels require the balancing of future free cash flows, and in turn, result in firm valuation changes.

Liquidity management requires that a sufficient balance of cash and other working capital assets - receivables and inventories – should be ensured[1]. If the level of liquid assets is not adequate, it enhances the company's operating risk – loss of liquidity. Maintenance of working capital assets generates costs, thus affecting the company's profitability. The problem of this paper is how liquidity can be combined with profitability.

If the level of liquid assets is too low, then a company may encounter problems with timely repayment of its liabilities, while discouraging clients by an excessively restrictive approach to recovery of receivables or shortages in the offered range of goods. Therefore, the level of liquid assets cannot be too low.

**Figure 1**

**Liquidity level vs. profitability**

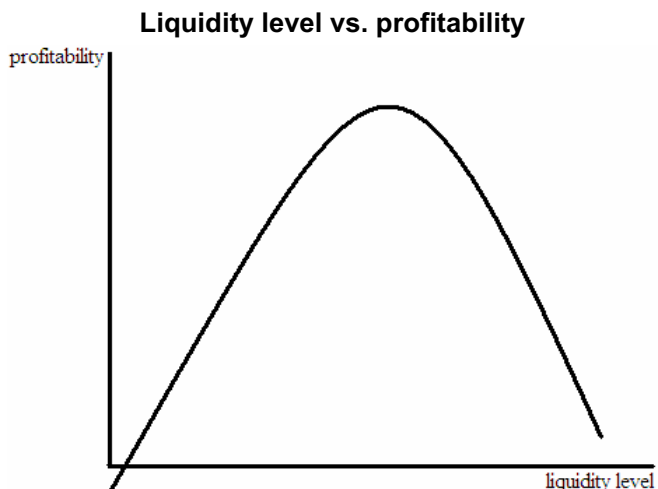

*Source: Michalski G., 2010, p. 25; Hundley G., 1996.*

At the same time (as we can see in Figure 1), surplus liquid assets may negatively affect the company's profitability. This is because upon exceeding the "necessary" level of liquid assets, their surpluses, when the market risk remains stable, become a source of ineffective utilisation of resources.

---

[1] Graham J.E., *Firm Value and Optimal Level of Liquidity*, Garland, New York 2001, pp. 4-6.

Along with an increased risk of the company's daily operations, you should increase the level of liquid assets to exceed the required levels as this will protect your company against negative consequences of unavailable liquid assets. It is possible to measure profitability of liquidity management decision in two ways. Firstly, it is possible to check how it affects the net profit and its relation to equity, total assets, or another item of assets. Secondly, it is possible to assess profitability in relation to value of the company.

Individual elements influencing liquidity management decisions affect the level of free cash flows to firm (*FCFF*) and thus the value of the company. Let us assume that the company is faced with a decision regarding the level of liquid assets. As we know, a higher debtors turnover ratio and inventory turnover ratio (resulting from a more liberal approach to granting a trade credit for the purchasers and offering a shorter turnaround on clients' orders) will be accompanied by more sales (larger cash revenues) but also higher costs.

**Figure 2**

**Liquid assets influence on value of the corporation**

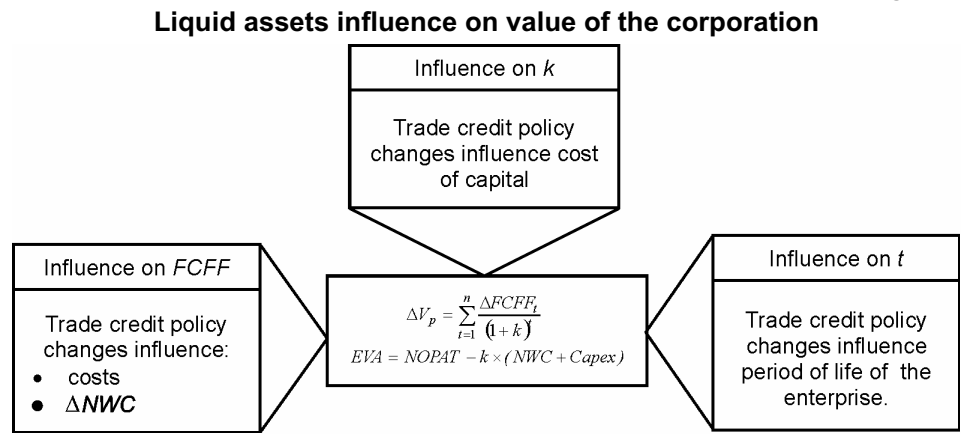

*where: FCFF = free cash flows to firm; ΔNWC = net working capital growth; k = cost of the capital financing the corporation; and t = the forecasted lifetime of the corporation and time to generate single FCFF.*
*Source: Michalski G., 2007, p. 45; Michalski G., 2008a, p. 84.*

Profitability measured by *ROE* indicates that "medium" liquidity level is optimal. Similar results will be achieved if estimating influence on the company's value (see Figure 2).

Again, the optimal variant was one that assumed a "medium" liquidity level as applying such level of liquidity ensures potentially the highest increase in the company's value measured by ΔV.

If the level of liquid assets is too low, it downsizes the sales thus discouraging clients with an overly restrictive trade credit policy. On the other hand, excessive exposure to liquid assets under the "high" level of liquid assets variant generated higher sales revenues than under the "medium" variant, but at the same time the positive result of increase in the sales volumes has been offset by high level of generated costs.

If the advantages of holding cash at a chosen level are greater than the influence of the alternative costs of holding cash, thereby increasing net working capital, then firm's value will also increase. The net working capital (current assets less current liabilities) results from lack of synchronization of the formal rising receipts and the real cash receipts from each sale. Net working capital also results from divergence during time of rising costs and time, from the real outflow of cash when a firm pays its accounts payable.

$$NWC = CA - CL = AAR + ZAP + G - AAP \qquad (1)$$

where: *NWC* = Net Working Capital, *CA* = Current Assets, *CL* = Current Liabilities, *AAR* = Accounts Receivables, *ZAP* = Inventory, *G* = Cash and Cash Equivalents, *AAP* = Accounts Payables.

When marking free cash flows, cash possession and increased net working capital is the direct result of amounts of cash allocated for investment in net working capital allocation. If an increase of net working capital is positive, then we allocate more money for net working capital purposes and thereby decrease future free cash flow. It is important to determine how changes in cash levels change a firm's value. Accordingly, we use equation, based on the premise that a firm's value is the sum of its discounted future free cash flows to the firm.

$$\Delta V_p = \sum_{t=1}^{n} \frac{\Delta FFCF_t}{(1+k)^t}, \qquad (2)$$

where: $\Delta V_p$ = Firm Value Growth, $\Delta FCFF_t$ = Future Free Cash Flow to Firm Growth in Period *t*, *k* = Discount Rate[2].

Future free cash flow we have as:

$$FCFF_t = (CR_t - FC_{WD} - VC_t - NCE) \times (1-T) + NCE - \Delta NWC_t - Capex_t \qquad (3)$$

where: $CR_t$ = Cash Revenues on Sales, $FC_{WD}$ = Fixed Costs, $VC_t$ = Variable Costs in Time *t*, *NCE* = Non Cash Expeses (i.e. Depreciation), *T* = Effective Tax Rate, *ΔNWC* = Net Working Capital Growth, *Capex* = Operational Investments Growth.

Changes in precautionary cash levels affect the net working capital levels and as well the level of operating costs of cash management in a firm. Companies invest in cash reserves for three basic reasons:

First, firms are guided by transactional and intentional motives resulting from the need to ensure sufficient capital to cover payments customarily made by the company. A firm retains transactional cash to ensure regular payments to vendors for its costs of materials and raw materials for production. As well, a firm retains intentional cash for tax, social insurance and other known non-transactional payment purposes.

Second, firms have precautionary motives to invest in cash reserves in order to protect the company from the potential negative consequences of risk, which are unexpected, negative cash balances that can occur as a result of delays in accounts receivable collection or delays in receiving other expected monies.

---

[2] To estimate changes in cash management, we accept discount rate equal to the average weighed cost of capital (WACC). Such changes and their results are strategic and long term, although they refer to cash and short term area decisions [T.S. Maness 1998, s. 62-63].

Third, companies have speculative motives[3] to retain cash reserves. Speculative cash makes it possible for the firm to use the positive part of the risk[4] equation to its benefit. Companies hold speculative cash to retain the possibility of purchasing assets at exceptionally attractive prices.

**Figure 3**
**Reasons for holding cash by companies and their relation to risk**

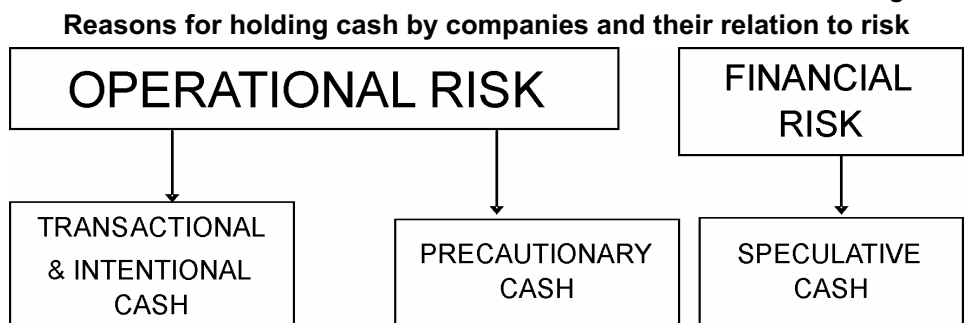

*Source: Michalski G., 2008b, p. 26.*

## 2. Value-Based Strategy in Working Capital Management

The issue discussed here attempts to address the question of which net working capital management strategy should be applied to bring the best results for a specific type of business. Financial decisions of a company always focus on selecting the anticipated level of benefits in conditions of risk and uncertainty. Decisions regarding net working capital management strategy, whether focused on assets (strategy of investing in the net working capital) or liabilities (strategy of financing the net working capital) affect free cash flows and the cost of capital financing the company. The principle of separating financial decisions from operating decisions, i.e. separating consequences of operations from changes in the capital structure, calls for a need to take the net working capital management decision first focusing on assets (it affects free cash flows to the company) and then on liabilities (it affects the structure and cost of capital used for financing the company).

Management of net working capital aimed at creation of value of the company. If the benefits of maintaining net working capital at the level determined by the company outweigh the negative influence of the alternative cost of such maintenance, then an increase in net worth of the company will be reported.

---

[3] [M.H. Miller 1966, s. 417-418].
[4] We define risk as the probability of obtaining a different effect than anticipated. Companies hold speculative cash to benefit from chance. Chance is the positive part of the risk equation, or the probability of obtaining an effect that is better than anticipated.

Interesting from our point of view, determined by the need to obtain the main objective of the company's financials management, is how a change in the net working capital level may impact the value of the company.

Net working capital is, most generally, the portion of current assets financed with permanent funds. The net working capital is a difference between current assets and current liabilities or a difference between permanent liabilities and permanent assets. It is a consequence of dichotomy between the formal origination of the sales revenue and the actual inflow of funds from recovery of receivables and different times when costs are originated and when the funds covering the liabilities are actually paid out.

When estimating free cash flows maintaining and increasing net working capital means that the funds earmarked for raising that capital are tied. If the increase is positive, it means ever higher exposure of funds, which reduces free cash flows for the corporation. An increase in production usually means the need to boost inventories, receivables, and cash assets. A portion of this increase will be most probably financed with current liabilities (which are also usually *automatically* up along with increased production volumes). The remaining part (indicated as an increase in net working capital) will need an alternative source of financing.

**Figure 4**

**Aggressive strategy**

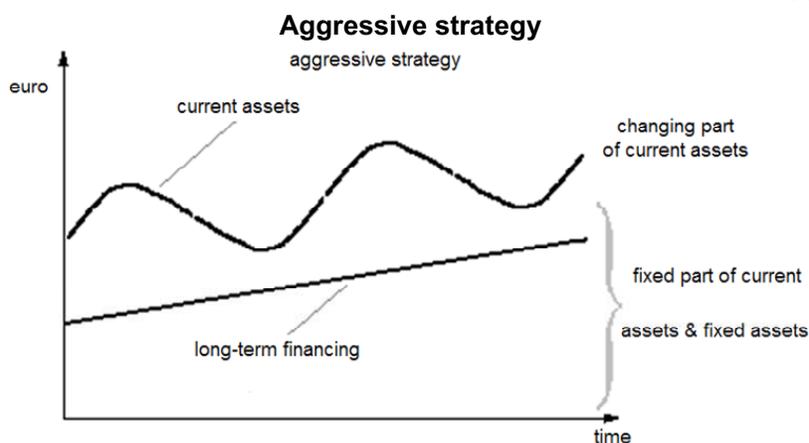

*Source: Michalski G. 2010, p. 158-160.*

Current asset financing policies are driven by the manner of financing current assets. Any changes to the selected current asset financing policy affect the cost of capital but do not impact the level of free cash flows. The company can choose one of the three policies:

a) an aggressive policy whereby a major portion of the company's fixed demand and the entirety of its volatile demand for financing current assets is satisfied with short-term financing.

b) a moderate policy aiming to adjust the period when financing is needed to the period when the company requires given assets. As a result of such approach, a fixed portion of current assets is financed with long-term funds, while the volatile portion of these assets is financed with short-term funds.

c) a conservative policy whereby both fixed and volatile levels of current assets are maintained with long-term financing.

The aggressive policy will most probably mean the highest increase in the net worth of the company. However, this result is not that obvious. This is because an increase in financing with an external short-term capital and a decrease in financing with an external long-term capital (namely shifting from conservative to aggressive policy of financing current assets) means enhanced risk level. Such increased risk level should be reflected in an increased cost of own capital. This stems from increased costs of financial difficulties.

The aggressive policy of financing current assets is the least favourable, considering an increased cost of own capital.

**Figure 5**

**Conservative strategy**

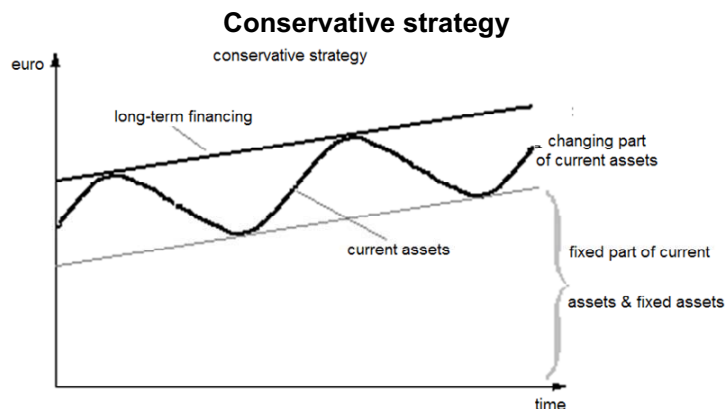

*Source: Michalski G. 2010, p. 158-160.*

Policies regarding investments in current assets are applied by the company as measures determining amounts and structure of current assets. There are three major policies available:

a) an aggressive policy whereby the level of tangible assets is minimised and a restrictive approach to merchant lending is applied. Minimising current assets results, on the one hand, in savings which later translate to higher free cash flows. On the other hand, insufficient level of current assets increases the operational risk. Too low inventories may interrupt the production and sales process. Insufficient level of receivables will most often lead to a restrictive merchant lending policy and, consequently, potentially lower sales revenue than in the case of a liberal merchant lending policy. Insufficient transactional cash levels may disrupt settlement of liabilities and as a result negatively affect the company's reputation.

b) a moderate policy whereby the level of current assets, and in particular inventories and cash, is held on an average level.

Planning Optimal from the Firm Value Creation Perspective

c) a conservative policy whereby a high level of current assets (and especially inventories and cash) is maintained at the company and ensuring a high level of receivables by using a liberal trade creditors recovery policy.

If the company aims at maximising ΔV, it should select the aggressive policy. However, similarly as in the preceding item, it is worth considering the relation between the risk increase and the cost of own capital (and probably also external capital). The more aggressive the current asset investment policy, the higher risk. Higher risk, on the other hand, should be accompanied by higher costs of own capital and probably also external capital.

Changes of the policy, from conservative to aggressive, cause an increase in the cost of capital financing the company's operations due to enhancement of risk. It is possible that in specific circumstances, the risk may drive the cost of capital to such a high degree that the aggressive policy will be unfavourable.

In the discussed examples, the company should select a conservative current asset financing strategy and an aggressive current asset investment policy.

The primary objective of financing the company's operations is to maximise the company's net worth. It can be estimated among others by totalling all the future free cash flows generated by the company, discounted with the cost of capital. Decisions regarding management of net working capital should also serve the purpose of achieving the primary objective, that is maximising the company's net worth. These decisions may impact both the level of free cash flows and the cost of capital used for financing the company's operations. The module discusses probably changes of the capital cost rate, resulting from changes in selection of the net working capital management policy and, consequently, the anticipated impact of such decisions on the company's net worth.

## 3. Value-Based Strategy in Cash Management

The most liquid current assets are cash balances. The purpose of cash management is to determine the level of cash resources at the company so that it increases the wealth of the company owners. In other words, the objective is to maintain such level of cash resources at the company that is optimal from the point of view of trade-off between the costs of maintaining cash balances against the costs of holding insufficient cash balances. The type and amount of these costs is partially driven by the particular financial policy applied by the company.

Based on observation of current inflows and outflows of the company, it may be noticed that there are four basic situations at the company in terms of operational cash flows:

1. when future inflows and outflows are foreseeable and inflows exceed outflows,
2. when future inflows and outflows are foreseeable and outflows exceed inflows,
3. when future inflows and outflows are foreseeable but it is impossible to determine which are in excess of which,
4. when future inflows and outflows are not foreseeable.

Depending on the type and volumes of inflows and outflows at the company, it is possible to select one of the four models of cash flow management. It is certainly not necessary for only one of the above situations to prevail at the company. The same business may have periods when inflows exceed outflows on a permanent basis, as well as periods when a reversed trend is noted or it is not possible to determine the trend. It is similar in case of projecting future inflows and outflows. It is possible that in some periods of time inflows and outflows can be projected without any major difficulty, while in other periods such projection is very hard or completely impossible.

Using information about future cash inflows and outflows, we are able to apply, for example, the Baumol model or the Beranek model. If we anticipate that cash inflows are greater than outflows, we are able to use the Beranek model [W. Beranek, 1963, also: F. C. Scherr, 1989, pp. 131-132] to determine cash flow management within a firm. On the other hand, if we predict that cash outflows are greater than inflows we use Baumol model [W. Baumol, 1952]. When we cannot forecast long-term cash flows, for a period longer than approximately 14 days, we are able to use the Stone model [B. Stone, 1972; T. W. Miller, 1996] to determine cash flow management. However, when we cannot predict future cash inflows and outflows at all, the Miller-Orr model[5] can be used to determine cash flow management.

According to the BAT model assumptions, the company receives both regular and periodic cash inflows, while it spends cash in an ongoing manner, at a fixed rate. At the time of receiving funds, the company earmarks a sufficient portion of these funds to cover its outflows. This is performed until the next inflow of cash. This model can be recommended in a situation when future inflows and outflows related to operations of the company can be foreseen and, at the same time, operational outflows exceed inflows. The BAT model comprises two types of assets: cash and (external) marketable securities, which generate profit in the form of interest during each period.

**Figure 6**

**The BAT model**

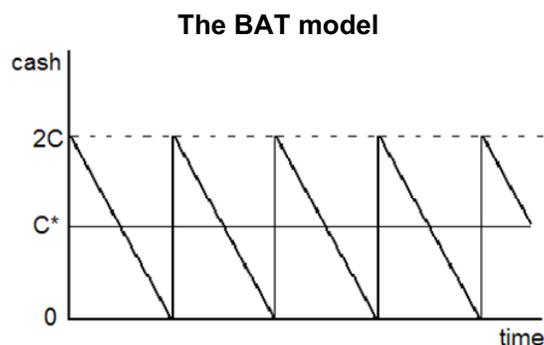

*Source: Beehler P.J. 1978, p. 191-200.*

The BAT model has been developed for two reasons: in order to specify the optimal cash balance at the company and to suggest how the company managers should proceed to ensure optimal cash management.

---

[5] M. H. Miller, 1984.

*Planning Optimal from the Firm Value Creation Perspective*

**Application of the model** is possible when we know with great probability the future inflows and outflows of the firm and we can predict that the outflows could be higher than inflows. **That model we could use** for example in the firm who has seasonal supply of the sources to production in the time of the collection of the inventories of the raw materials needed in the time of production.

The company which decides to follow recommendations regarding cash management, arising from the BAT model, determines an optimal cash level $C^*_{bau}$.

It stems from the BAT model that when cash is spent, the company should secure cash from non-operational sources of cash. Most often, this means that it should sell (external) securities, close the held deposit, and/or raise a short-term loan. The total amount of raised funds should be in each event twice as high as an average cash balance. The ratio of the total demand for cash in a given period and one transfer, provides information on how many such operations must be performed during the year. It is clear that if conditions, which enable application of the BAT model, have existed at the company for less than one year, then shorter periods should be taken into account.

Similar to the BAT model is Beranek Model. That model is for another situation. **Application of the Beranek model** is possible when we know with great probability, the future inflows and outflows of the firm, and we can predict that the inflows could be higher than outflows. **Beranek model we could use** for example in the firm who has seasonal supply of the sources to production in the time after the collection of the inventories of the raw materials needed in the time of production.

**Figure 7**

**The Beranek model**

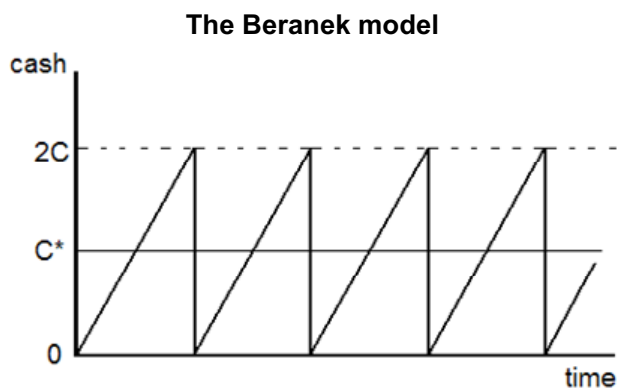

*Source: Beehler P.J. 1978, p. 191-200.*

The basic assumption of the Miller-Orr model is that changes in cash balance at the company are unforeseeable. The company managers react automatically when cash balance equals either the upper or lower level. This model is presented in the figure 8.

**Application of the Miller-Orr model** is possible when we do not know the future inflows and outflows of the firm at all, and we can only predict the general level of outflows, but we cannot predict which will be higher, inflows or outflows. **That model**

*Institute of Economic Forecasting*

**we could use** for example in the firm who has dynamic situation as seller of their production, without the guarantee if and when the goods will be sold.

**Figure 8**

**The Miller-Orr Model**

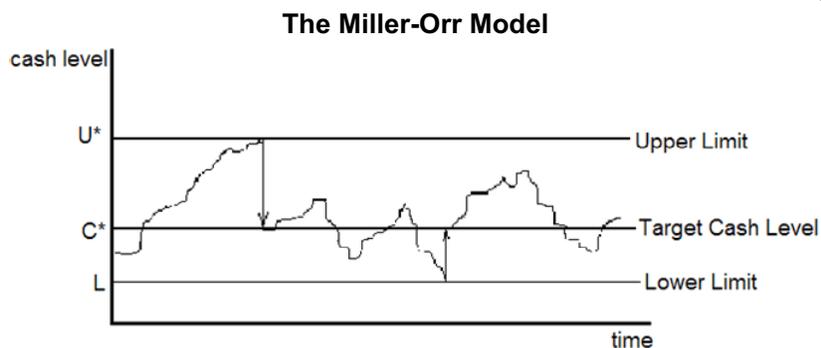

*Source: Beehler P.J. 1978, p. 191-193.*

Reacting to the situation when the cash balance at the company reaches the upper or lower limit, the management board buys or sells (external) short-term securities, opens or closes short-term deposits and/or repays or raises a short-term loan in order to restore the target cash balance $C_{mo}^*$.

This model is used traditionally in such a manner that the management board of the company first specifies the lower limit of cash $L$ that it finds acceptable. This value is specified subjectively based on experience of the company managers. As in a sense it is a minimum level of cash balance, it depends on such factors as availability of the company's access to external financing sources. If in the opinion of the management board members this access is easy and relatively inexpensive, liquidity at the company is lower and $L$ can be set on a relatively low level.

The Miller-Orr model assumes that the target cash balance $C^*$ depends on the (alternative) costs of holding funds, costs of cash shortages (transfer) and variants of cash flows during the considered period (this period must equal the period for which an interest rate has been set). The level of variance of cash flows during the analysed period is best determined based on historic data.

The target cash balance according to the Miller-Orr model is calculated based on the formula for $C^*_{mo}$:

In this model, after setting the target cash balance $C^*_{mo}$ the upper limit $U^*$ is determined as a difference between triple target cash balance and double lower control limit.

The Stone model is a modification of the Miller-Orr model for the conditions when the company can forecast cash inflows and outflows in a few-day perspective. Similarly to the Miller-Orr model, it takes into account control limits and surpassing these limits is a signal for reaction. In case of the Stone model, however, there are two types of limits, external and internal, but the main difference is that in case of the Stone model, such signal does not mean an automatic correction of cash balance as in the Miller-Orr model.

**Figure 9**

**The Stone Model**

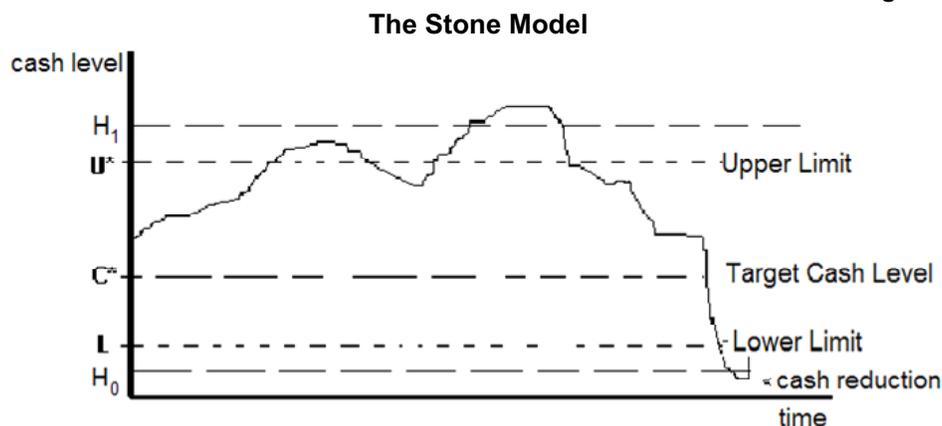

*Source: Beehler P.J. 1978, p. 191-200.*

If the cash balance exceeds the upper external limit $H_1$ or the lower external limit $H_0$, the management board analyses future cash inflows by projecting future cash balance by calculating the S level.

If the *S* level (determining the cash balance after *n* days from the moment of surpassing either of the external control limits) continues to surpass any of the internal limits, the management board should prevent variations from the target balance by purchase or disposal of securities in the amount sufficient for the cash balance at the company to be restored to its optimal level $C_s^*$.

This model is presented in Figure 9. It shows that the cash balance has been growing as from the beginning of the analysed period. At some point, it exceeded the upper internal limit $U^*$. Then it exceeded the external control limit $H_1$. At the time of exceeding the external control limit, the management board of the company forecast future inflows and outflows. As the forecast indicated that the cash balance would continue to exceed the internal control limit (the grey line), the management board decided to adjust this level to the anticipated $C^*$. After the appropriate adjustment, the cash balance started to decrease after a few days and it surpassed the lower external control limit. Another forecast was prepared and it turned out that for several days the cash balance would remain below the lower internal control limit. Therefore, the cash balance was reduced down to $C^*_S$.

**Application of the Stone model** is possible when we generally do not know the future inflows and outflows of the firm but we can partially predict them for short time. We can also predict the general level of outflows, but for the time after the short time we have our little knowledge about inflows and outflows, we cannot predict which will be higher, inflows or outflows. **The Stone model we could use** for example in the firm who has dynamic situation as seller of their production, without the guarantee if and when the goods will be sold, but they know with little advance of few days, what will be sold or bought.

## 4. Cash Balance Forecasting

Maintaining the appropriate cash balance requires not only ongoing monitoring of the currently held current assets and liabilities that mature in the forthcoming future, but also those that should be anticipated in the future. Therefore, it is necessary to plan future cash inflows and outflows[6].

Cash forecast is performed based on cash budget. This tool contains a forecast of recovered receivables, expenditure on inventories and repayment of liabilities. It provides information about the cash balance, as cash balance is a result of inflows from sales (payment of receivables) and outflows due to purchase of materials and other costs of the company.

Cash budget is most often prepared several periods in advance subject to the company's information capabilities and needs. The most popular version of a cash budget is one prepared for six monthly periods. However, there are no reasons why cash budgets should not be prepared for six weeks or six biweekly periods. In any case, the rolling wave planning is used, which requires that subsequent periods be added to the budget on an ongoing and regular basis so that at any one time the company has a forecast for the fixed number of forecast periods (namely, if the budget is prepared for 8 biweekly periods, then it should be adequately extended when required so that a sixteen-week budget is available at any time). This requirement ensures for the budget to be constantly valid and applicable. Six-month budgets are most frequently prepared based on monthly time bands. For some companies it is absolutely necessary to determine inflows and outflows for individual weeks and sometimes even days. The more detailed the control of cash inflows and outflows, the more probable the precise and correct control of cash flow levels. When developing a cash budget, it is a matter of top priority to hold a forecast of the company's sales revenues. Preparing such forecast is the primary and at the same time the hardest task. Next, the demand arising from the held fixed assets and inventories, resulting from production of goods for sale, is forecast. This information is combined with information on delays in recovery of receivables. Also tax due dates, interest due dates, and other factors are taken into account.

## 5. Precautionary Cash Management - Safety Stock Approach

Current models for determining cash management, for example Baumol, Beranek, Miller-Orr or Stone models, assign no minimal cash level, and are based on the manager's intuition. In addition, these models are based inventory managements models. In this study, we address the potential for adaptation of these methods of determining safety stock to determine minimal cash levels in the firm. Safety stock is a result of information about the risk of inventories. To calculate safety stock we use Equation 4 [M. Piotrowska, 1997, p. 57]:

---

[6] *Moir, L.,* Managing corporate liquidity*, Woodhead, Glenlake/Amacom, New York 1999, pp. 11-39.*

*Planning Optimal from the Firm Value Creation Perspective*

$$Z_b = \sqrt{-2 \times s^2 \times \ln \frac{C \times Q \times s \times v \times \sqrt{2\Pi}}{P \times K_{bz}}} \qquad (4)$$

where: $z_b$ = Safety Stock, $C$ = Cost of Inventories (in percentage), $Q$ = One Order Quantity, $v$ = Cost of Inventories (Price), $P$ = Yearly Demand for Inventories, $s$ = Standard Deviation of Inventory Spending, $K_{bz}$ = Cost of Inventories Lack.

It is also possible to apply the following equation to determine minimal cash level [G. Michalski 2006]:

$$LCL = \sqrt{-2 \times s^2 \times \ln \frac{k \times G^* \times s \times \sqrt{2\Pi}}{P \times K_{bsp}}} \qquad (5)$$

where: $LCL$ = Low Cach Level (Precautionary Cash Level), $k$ = Cost of Capital, $G^*$ = Average Size of One Cash Transfer[7] which are the basis of standard deviation calculation, $P$ = the Sum of all Cash Inflows and Outflows in the Period, $s$ = Standard Deviation of Daily Net Cash Inflows/Outflows, $K_{bsp}$ = Cost of Cash Lack.

Part of the information necessary to determine *LCL*, still requires the manager's intuition. For example, costs of lack of cash, contains not only costs known from accountant records, but also other costs, such as opportunity costs. Precautionary cash reserves are, first of all the result of anxieties before negative results of risk. Its measure is the standard deviation.

**Case 1.** Suppose, that managers of the hypothetical firm X, value the cost of the lack of cash as 5000. The managers know from historical data of their hypothetical firm, that the day's standard deviation of cash inflows/outflows is 35,466 monthly. Average single cash inflow/outflow is 27,250. The monthly sum all cash inflow/outflow is: 817,477. The alternative cost of capital is 18%.

For the firm X, the precautionary cash level is:

$$LCL_1 = \sqrt{-2 \times 35\,466^2 \times \ln \frac{\frac{0,18}{360} \times 27\,250 \times 35\,466 \times \sqrt{2\Pi}}{817\,477 \times 5\,000}} = 142\,961.42$$

When cash outflows and inflows volatility is 0, precautionary cash balance is also 0:

$$LCL_0 = 0$$

Then we can estimate net working capital growth:

$$\Delta NWC = LCL_1 - LCL_0 = 142\,961{,}42 = -\Delta CF_{t=0}$$

The standard deviation is 35,466 and tax rate is 20%. So, we can estimate yearly alternative cost precautionary cash reserves and the influence on the value of the firm:

---

[7] In Beranek model and Baumol models, G* is twice optimal cash level. In Stone and Miller-Orr models, the average transfer G* is assigned from real historic data or from its anticipation.

*Institute of Economic Forecasting*

$$\Delta TCC = \Delta NWC \times k = 142\,961.42 \times 0.18 = 25{,}733 = \frac{-\Delta CF_{t=1\ldots\infty}}{(1-T)};$$

$$\Delta V = \Delta CF_{t=0} + \frac{(\Delta CF_{t=1\ldots\infty}) \times (1-T)}{k} = -142{,}961.42 + \frac{-25{,}733 \times 0.8}{0.18} = -257{,}330$$

As demonstrated in order for the precautionary cash balance to remain level, with the standard deviation equal to 35,466; a decrease in the firm's value of 257,330 results.

## 6. Speculative Cash Balance Management - Option Approach

All firms do not necessarily hold speculative cash balances. Speculative cash is held in order to utilize the positive part of the risk equation. Firms want to retain opportunities that result from price volatility. For example, in the ordinary practice of Polish firms, we see that speculative cash balances can be useful to benefit from transactions in foreign exchanges. It can be profitable for firms to purchase necessary products or services in foreign exchange at prices cheaper than its average purchase price. Such purchase is possible if the firm maintains speculative cash balances. Speculative cash balances give the firm the ability to use of their purchasing power any time. Such cash superiority over other assets shows option value of speculative cash balances[8].

**Case 2.** Based on [Beck S.E., 1993]. Suppose, that managers of the hypothetical firm X, based on their historical data, could choose from one of two possibilities:

- they could to invest in the firm activity, for example, he can purchase in foreign exchange,

or

- they could decide to hold cash (national currency).

The management makes the decision between these two possibilities at least once every day. The purchase of foreign exchange and its use in the operating activity of a firm makes other cash resources inaccessible for continued speculation. If the management chooses to hold cash, he still has the possibility to purchase foreign exchange. Yet, foreign exchange price changes from day to day. The daily standard deviation of the foreign exchange price is 4%. This means that the foreign exchange price today is 1.00 PLN. The next day the foreign exchange price can be 1.04 PLN with the probability 0.5; or 0.96 PLN with the probability 0.5.

Suppose that next, the foreign exchange price meets its long-term value of 1.00 PLN. If on the first day, the management decides to hold cash, and the next day's foreign exchange price falls to the level of 0.96 PLN (lower than its expected value), the firm's expected income will be 0.04 PLN. On the other hand, if the foreign exchange price reaches the level of 1.04 PLN (above its expected value), then the firm won't purchase foreign exchange, and his expected income will be 0 PLN. So, if an entrepreneur has cash for 10,000 foreign exchange units, his expected value of the benefit of holding in national currency (in cash) by one day, will be:

$$\text{E(benefit)} = \sum_{i=1}^{n} \text{benefit} \times p_i = \frac{0.04 \text{PLN} \times 10,000}{1.0005} \times 0.5 + 0 \text{PLN} \times 0.5 \approx 199.90 \text{ PLN}$$

---

[8] Beck, S.E., 1993. Cash we can compare to American option without *expiration date*. Other near to cash assets can be compared to European option, see: [J. Ingersoll, 1992, pp. 5-6, and S. E. Beck, 1993]. The right to faster acquisition has a value, and such value gives base to have speculative cash balances. Costs of expectation on realization of other options can cause loss that is not recovered by future earnings from these (less liquid than cash) assets, see [S. E. Beck, 2005].

The daily alternative cost of capital financing for the firm is:

$$\frac{18\%}{360} = 0.05\%$$

Therefore, we can also express it for 10,000 foreign exchange units:

$$0.05\% \times 10,000 = 5 \text{ PLN}.$$

This means that the expected benefit is 199.9 PLN. This demonstrates the basis for holding speculative cash balances in a firm. Of course, the size of speculative cash balances should be an effect of the firm's customary activities and its real operational needs. The legitimacy of holding speculative cash balances increases here together with the increase of volatility of foreign exchange pricing (or volatility of the price of any other assets necessary to the firm) and grows smaller together with the height of the alternative costs of capital financing for the firm.

## 7. Conclusions

Liquid assets management decisions are very complex. On the one hand, when too much money is tied up in working capital, the business face higher costs of managing liquid assets with additional high alternative costs. On the other hand, the higher liquidity assets policy could help enlarge income from sales. Firms hold cash for a variety of different reasons. Generally, cash balances held in a firm can be called considered, precautionary, speculative, transactional and intentional. The first are the result of management anxieties. Managers fear the negative part of the risk and hold cash to hedge against it. Second, cash balances are held to use chances that are created by the positive part of the risk equation. Next, cash balances are the result of the operating needs of the firm. In this article, we analyze the relation between these types of cash balances and risk. This article also contains propositions for marking levels of precautionary cash balances and speculative cash balances. Application of these propositions should help managers to make better decisions to maximize the value of a firm.

## References

Baumol, W. (1952), "The Transactions Demand for Cash: An Inventory Theoretic Approach", *Quarterly Journal of Economics*, pp. 545-556.